\documentclass[
 reprint,
 amsmath,amssymb,
 aps,
]{revtex4-1}

\usepackage{amsmath,bm,morefloats}  
\usepackage{amsfonts} 
\usepackage{graphicx} 
\usepackage{CJK}

\usepackage[displaymath, mathlines]{lineno}

\begin{document}

\begin{CJK*}{UTF8}{}
 \CJKfamily{mj}

\title{Imperfect Energy Conservation of the Free-Electron Laser over the Timestep of Simulations}

\author{Jeongwan Park}

\email{jwpark@hawaii.edu} 
\affiliation{Department of Physics, University of Hawaii at Manoa, Honolulu, HI 96822}



\date{\today}

\begin{abstract}
 Along the development of free-electron laser operating at the wavelength of X-ray, the importance of investigation on the radiation has increased. A theoretical simulation is an essential tool for studying existing and proposed experiments. The available simulations preserve net energy conservation over the timestep, upholding causality between the electrons' power and the radiated energy flux. However, according to Wheeler-Feynman time-symmetric theory, the timestep is too short to ensure this. Therefore, the time evolution of electrons and radiation field predicted by the simulations should contradict the correct physics, regardless of whether the stimulated emission dominates over the spontaneous emission or not.
\end{abstract}

\maketitle 
\end{CJK*}


\section{Introduction}Along the development of free-electron laser (FEL) operating at the wavelength of X-ray, the importance of FEL has increased. The X-ray radiation generated by available FELs is intense but only partially coherent \cite{madey0}. Therefore, much work is being done to improve coherence \cite{pellegrini}. Consequently, a detailed analysis of the physics of existing or proposed FEL has become more significant. In such an analysis, as analytic predictions cannot be easily made, theoretical simulation has become an integral tool. Therefore, apprehending any possible physical limitations of the simulations is vital for the improvement of the FEL.

Within the theoretical simulations, the FEL equations are numerically integrated to compute the time evolution of electrons and radiation field. The FEL equations are composed of a set of three equations: one equation for the electron energy, one for the field amplitude, and a third for the electron phase of the ponderomotive potential \cite{eric}\cite{elias}. The FEL equations base on an approximation applied to the field of classical electrodynamics (CED) that builds on the retarded only formulation and do not infer any singularity of the field, which overlooks the issue of the singularity of CED's field. The simulations' information of electrons and radiation field is updated every timestep, the time it takes for a radiation field slice to slip over an electron beam slice. The update is in accordance with preservation of net energy conservation over the timestep, upholding causality between the electrons' power and the radiated energy flux, which is dictated by the FEL equations regardless of the number of radiating electrons \cite{kim}. It should be worthwhile investigating whether the dictation is physically correct because if that is not true, the simulations can contradict the actual physics.

To check the validity of net energy conservation over the timestep, according to the integral form of Poyntings' theorem, the time integrations of electrons' power and radiated energy flux, which are causally connected, over the timestep should be compared. And the electrons' power should be computed from the actual field at the electrons, not from an approximated field. Hence, the FEL equations' implication of net energy conservation over the timestep cannot mean that it is indeed preserved. However, whether it is preserved has not been investigated in detail. Therefore, in this paper, the validity of net energy conservation over the timestep is investigated, using the actual field at the electrons.

\section{Energy conservation investigated by Wheeler-Feynman time-symmetric theory}
In CED the time integrations of electrons' power and radiated energy flux over the timestep cannot be compared as the electrons' power cannot be computed due to the indeterminateness of the field at the electrons. Therefore, it may be attempted to calculate the electrons' power by assigning a portion of radiated energy flux to each electron in an \textit{ad hoc} manner \cite{elias}. However, in such attempts, electrons' power is calculated using CED's radiated energy flux and an assumption of energy conservation. Therefore, the calculated electrons' power cannot be used to investigate the validity of energy conservation. Using CED, the validity of net energy conservation over the timestep cannot be investigated.

To better study the field at the charged particles radiating into free-space, overcoming CED's issue of the indeterminate field at the charged particles, Dirac found a non-retarded, finite, and definite expression for the field at the charged particles \cite{dirac}. However, the physical origin of Dirac's field is not clear \cite{feynman}. On the other hand, an alternative theory of CED, Wheeler-Feynman time-symmetric theory of complete absorbing system (CWFT) \cite{feynman}, which bases on the first principle of time-symmetric interaction between the charged particles and the absorber of complete absorbance, infers that at the charged particles the field is the same as Dirac's field. Therefore, it is well worth considering computing the electrons' power using CWFT to investigate the validity of net energy conservation over the timestep.

Some aspects of CWFT remain controversial and it is conceptually challenging, but it is not aggressive to CED \cite{madey0}; at large distance from the electrons radiating into free-space, the radiated energy flux in CED and CWFT become the same \cite{feynman}. Furthermore, the CWFT's time integration of electrons' power over a period is shown to match the time integration of CED's radiated energy flux at large distance from the electrons, for the case of a periodically oscillating electron \cite{feynman}, and the case of two synchronously periodically oscillating non-relativistic electrons spaced by distances of the order of a wavelength \cite{madey0}. Therefore, CWFT can be considered as a solid alternative theory of CED, which infers a definite field at the electrons. Hence, this paper's investigation of the validity of FEL's net energy conservation over the timestep is done using CWFT.

To simplify the analysis, which is done in the FEL electrons' comoving frame, the radiation into free-space of a periodically oscillating electron or two periodically oscillating electrons is investigated.
As the timestep of simulations should be less than a period because the spacing between the electrons can be smaller than a radiation wavelength, the validity of net energy conservation over time less than a period is studied. By studying two cases of the different number of the electrons, the trend of the validity of net energy conservation depending on the number of the electrons can be studied, from which an indicator of the validity of the more complicated FEL's net energy conservation over the timestep can be obtained. The single electron's radiation mimics the case of spontaneous emission whereas the two electrons' radiation mimics the case when both the spontaneous and stimulated emissions are present. Electrons are assumed to be non-relativistic (in the FEL electrons' comoving frame, electrons can be approximated to be non-relativistic \cite{fel}) and point-like in the analysis. And assumed are that the radiated energy flux originates from the causal origin (the electrons' power) and that there is no singularity of the field. It is found that net energy conservation over time less than a period is not guaranteed to be preserved.

\section{Energy conservation for single electron's radiation}To compute the radiated energy flux and electron's power, the field needs to be obtained first. As the electron is in the non-relativistic limit, \(\frac{U_0}{c\omega}\) [\(\mathbf{r}_s(t)=z_0 e^{-i\omega t}\hat{z}\) is the electron's position, and \(\mathbf{U}(t)\equiv \ddot{\mathbf{r}}_s(t)=-U_0e^{-i\omega t}\hat{z}\), where \(z_0\) and \(U_0\) are non-negative] is in the limit of \(\frac{U_0}{c\omega}\rightarrow 0\). The field at \(\mathbf{r}\) is given as the following at \(\mathbf{r}\) of order lower than \(\mathcal{O}\Big(\frac{U_0}{c\omega}\Big)\):

\begin{equation}\label{13}
\mathbf{E}_{abs}(\mathbf{r},t)+\frac{1}{2}\{\mathbf{E}_{ret}(\mathbf{r},t)+\mathbf{E}_{adv}(\mathbf{r},t)\},
\end{equation}
and at \(\mathbf{r}\) of order higher than \(\mathcal{O}\Big(\frac{U_0}{c\omega}\Big)^0\)the field is the following:

\begin{equation}\label{13-2}
    \mathbf{E}_{abs}(\mathbf{r},t).
\end{equation}
The distinction criteria between order of \(\mathbf{r}\) corresponding to Eqs. (\ref{13}) and (\ref{13-2}) is whether the order converges to zero in the limit of \(\frac{U_0}{c\omega}\rightarrow 0\) or not. \(\mathbf{E}_{ret}(\mathbf{r},t)\) and \(\mathbf{E}_{adv}(\mathbf{r},t)\) are the retarded and advanced field, respectively. \(\mathbf{E}_{abs}(\mathbf{r},t)\) lying within the plane \((\mathbf{r},\mathbf{r}_s)\) is the advanced field of the absorber. From the derivation of \(\mathbf{E}_{abs,\parallel}(\mathbf{r},t)\) [portion of \(\mathbf{E}_{abs}(\mathbf{r},t)\) parallel to \(\mathbf{U}(t)\)] in CWFT, \(\mathbf{E}_{abs,\perp}\) [portion of \(\mathbf{E}_{abs}(\mathbf{r},t)\) perpendicular to \(\mathbf{U}(t)\)] can be computed. Then, from the expressions of \(\mathbf{E}_{ret}(\mathbf{r},t)\), \(\mathbf{E}_{adv}(\mathbf{r},t)\), and \(\mathbf{E}_{abs}(\mathbf{r},t)\) [Eqs. (\ref{8}), (\ref{23}), and (\ref{17}), respectively], the following relation can be obtained:

\begin{equation}\label{eabs}
   \mathbf{E}_{abs}(\mathbf{r},t) =\frac{1}{2}\{\mathbf{E}_{ret}(\mathbf{r},t)-\mathbf{E}_{adv}(\mathbf{r},t)\}.
\end{equation}
Therefore, according to Eqn. (\ref{13}), the field of CED and CWFT are the same at \(\mathbf{r}\) of order lower than \(\mathcal{O}\Big(\frac{U_0}{c\omega}\Big)\).

From the obtained field, the electron's power can be computed as the following:
\begin{equation}\label{ew}
\begin{split}
    W(t)&\equiv -q\lim\limits_{\mathbf{r}\rightarrow \mathbf{r}_s}[\mathbf{E}_{abs}(\mathbf{r})]\cdot Re[\dot{\mathbf{r}}_s(t)]=\frac{q^2U_0^2}{6\pi \epsilon_0 c^3}\sin^2(\omega t).
\end{split}
\end{equation}
 The radiated energy flux over the boundary of a sphere of radius \(R\rightarrow \infty\) which is centered at the origin, \(\lim\limits_{R\rightarrow \infty}P(t,R)\), becomes the following [its detailed derivation is given in Eqn. (\ref{ef0})]:
\begin{equation}\label{pp1}
\lim\limits_{R\rightarrow \infty}P(t,R)=\lim\limits_{R\rightarrow \infty}\frac{q^2U_0^2}{6\pi \epsilon_0 c^3}\cos^2\{\omega(t-\frac{R}{c})\}.
\end{equation}


 To compare the time integrations of radiated energy flux and electron's power, the time difference between the radiated energy flux and electron's power which link them causally should be found. As the distance between the electron and the infinite sized sphere's boundary is \(\lim\limits_{R\rightarrow \infty}R\) [the displacement of electron's oscillation, which is of order of \(\mathcal{O}\Big(\frac{U_0}{c\omega}\Big)^2\), can be neglected], the electron can be considered to influence \(\lim\limits_{R \rightarrow \infty}P(t,R)\) causally, with the time retardation of \(\lim\limits_{R\rightarrow \infty}\frac{R}{c}\). Therefore, if the instantaneous energy conservation is preserved, \(\lim\limits_{R\rightarrow \infty} P(t+\frac{R}{c},R)\) should be equal to \(W(t)\). Consequently, a measure of the imperfection of net energy conservation over time of \(\Delta\) that is of order lower than \(\mathcal{O}\Big(\frac{U_0}{c\omega}\Big)^2\), \(I(\Delta)\), can be defined as the relative ratio between energies originating from two portions of \(W(t)\), as the following:

\begin{equation}\begin{split}\label{err}
&I(\Delta)\equiv\frac{\int_{t_0}^{t_0+\Delta}\lim\limits_{R\rightarrow \infty}\frac{1}{2}\{W(t)-P(t+\frac{R}{c},R)\}dt}{\int_{t_0}^{t_0+\Delta}\lim\limits_{R\rightarrow \infty}\frac{1}{2}\{W(t)+P(t+\frac{R}{c},R)\}dt}
\\&=-\frac{\sin (\omega \Delta)\cos \{\omega (2t_0+\Delta)\}}{\omega \Delta}.
\end{split}\end{equation}
Because the phase of oscillations of simulated FEL's electrons can be specified over the timestep, \(t_0\) is of order of a period, \(\mathcal{O}\Big(\frac{U_0}{c\omega}\Big)\). 
As \(I(\Delta)\) deviates more from zero, net energy conservation over \(\Delta\) is less perfectly preserved. \(I(\Delta)\) is zero not for every \(\Delta\).

\section{Energy conservation for two electrons' radiation}
\begin{figure}[!ht]
        \centering
        \includegraphics[height=3cm]{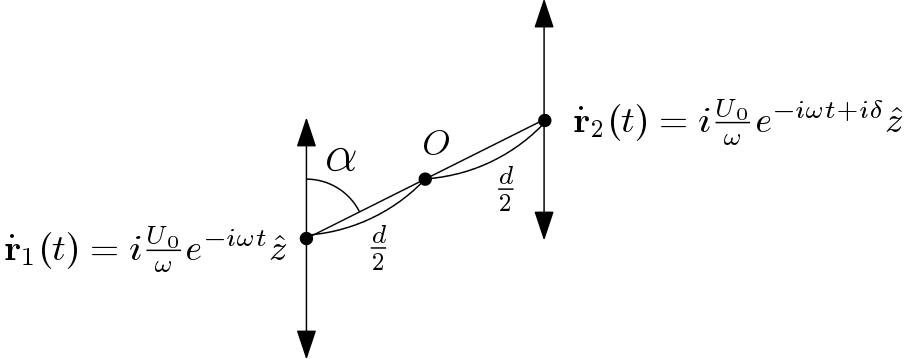}
        \caption{Two periodically oscillating electrons that radiate into free-space.}
        \label{2c}
    \end{figure}

In the case of radiation of two electrons named 1 and 2 as shown in Fig. \ref{2c} [\(d\), the spacing between the two electrons, is of order lower than \(\mathcal{O}\Big(\frac{U_0}{c\omega}\Big)^2\)], for \(d\) of order lower than \(\mathcal{O}\Big(\frac{U_0}{c\omega}\Big)\), \(\mathbf{E}^{(i)}(t)\), the field at the \(i\)th electron, is the following (the index \(i\) in the subscript of the field represents that the field is created because of the \(i\)th electron):

\begin{equation}\begin{split}\label{e1s2}
    &\mathbf{E}^{(1)}(t)=\lim_{\mathbf{r}\rightarrow \mathbf{r}_1}\{\mathbf{E}_{abs,1}(\mathbf{r},t)+\mathbf{E}_{ret,2}(\mathbf{r},t)\},
    \\&\mathbf{E}^{(2)}(t)=\lim_{\mathbf{r}\rightarrow \mathbf{r}_2}\{\mathbf{E}_{abs,2}(\mathbf{r},t)+\mathbf{E}_{ret,1}(\mathbf{r},t)\}.
    \end{split}\end{equation}
 For \(d\) of order of a wavelength, \(\mathcal{O}\Big(\frac{U_0}{c\omega}\Big)\), the field at the electrons becomes the following:

\begin{equation}\begin{split}\label{p1}
&\mathbf{E}^{(1)}(t)=\lim\limits_{\mathbf{r}\rightarrow \mathbf{r}_1}\{\mathbf{E}_{abs,1}(\mathbf{r},t)+\mathbf{E}_{abs,2}(\mathbf{r},t)\},
\\&\mathbf{E}^{(2)}(t)=\lim\limits_{\mathbf{r}\rightarrow \mathbf{r}_2}\{\mathbf{E}_{abs,1}(\mathbf{r},t)+\mathbf{E}_{abs,2}(\mathbf{r},t)\}.
\end{split}\end{equation}

From the obtained field, electrons' power can be computed as the following:

\begin{equation}\begin{split}\label{forcem}
    &W_{two}(t,\alpha,\delta,d)\\&=-q\Big[Re[\mathbf{E}^{(1)}]\cdot Re[\dot{\mathbf{r}}_1]+Re[\mathbf{E}^{(2)}]\cdot Re[\dot{\mathbf{r}}_2]\Big],
\end{split}
\end{equation}
which is calculated in Eqn. (\ref{wf}) for \(d\) of order lower than \(\mathcal{O}\Big(\frac{U_0}{c\omega}\Big)\), and in Eqn. (\ref{l5}) for \(d\) of order of \(\mathcal{O}\Big(\frac{U_0}{c\omega}\Big)\) (\(\alpha\) and \(\delta\) are defined in Fig. \ref{2c}). The radiated energy flux over the boundary of a sphere of radius \(R \rightarrow \infty\) (order of \(d\) is higher than that of \(R\), in terms of \(\frac{U_0}{c\omega}\)) centered at \(O\) (the origin) of Fig. \ref{2c}, \( \lim\limits_{R\rightarrow \infty}P_{two}(t,R,\alpha,\delta,d)\), is calculated as the following [its detailed derivation is presented in Eqs. (\ref{n00}) - (\ref{ef})]:

\begin{equation}\label{nn1}
\begin{split}
    &\lim\limits_{R\rightarrow \infty}P_{two}(t,R,\alpha,\delta,d)=\\&\lim\limits_{R\rightarrow \infty}\frac{q^2 U_0^2}{3\pi \epsilon_0 c^3} \cos^2 \{\omega(t-\frac{R}{c})-\frac{\delta}{2}\}\{1+3 X(d,\alpha)\cos \delta\},
     \end{split}
\end{equation}
where \(X(d,\alpha)\) is given in Eqn. (\ref{aas}).

  Similar to the single electron's radiation, the time difference between radiated energy flux and electrons' power which connect them causally should be obtained, to compare the time integrations of radiated energy flux and electrons' power. Comparing \(\lim\limits_{R\rightarrow \infty}P_{two}(t,R,\alpha,\delta,d)\) to \(\lim\limits_{R\rightarrow \infty}P(t,R)\), it can be noted that if the two electrons are replaced by an equivalent electron centered at the origin with acceleration of \(\ddot{\mathbf{r}}'_s=-\sqrt{2\{1+3 X(d,\alpha)\cos\delta\}}U_0e^{-i(\omega t -\frac{\delta}{2})}\hat{z}\), the radiated energy flux created by the equivalent electron, \(\lim\limits_{R\rightarrow \infty}P'(t,R,\alpha,\delta,d)\), is equal to \(\lim\limits_{R\rightarrow \infty}P_{two}(t,R,\alpha,\delta,d)\). Moreover, according to \(W(t)\) and \(W_{two}(t,\alpha,\delta,d)\), if \(\delta=2n\pi\) (\(n\) is an integer), the equivalent electron's power, \(W'(t,\alpha,\delta=2n\pi,d)\), is equal to \(W_{two}(t,\alpha,\delta=2n\pi,d)\). For the equivalent electron's radiation, the causality dictates that \(W'(t,\alpha,\delta=2n\pi,d)\) is the cause of \(\lim\limits_{R\rightarrow \infty}P'(t+\frac{R}{c},R,\alpha,\delta=2n\pi,d)\). This can infer that \(W_{two}(t,\alpha,\delta=2n\pi,d)\) is the effective cause of \(\lim\limits_{R\rightarrow \infty}P_{two}(t+\frac{R}{c},R,\alpha,\delta=2n\pi,d)\) for the two electrons' radiation, although the distance between an electron and a point within the sphere's boundary is not fixed. Therefore, if the instantaneous energy conservation is preserved for the radiation of two electrons, \( W_{two}(t,\alpha,\delta=2n\pi,d)\) should be equal to \(\lim\limits_{R\rightarrow \infty}P_{two}(t+\frac{R}{c},R,\alpha,\delta=2n\pi,d)\).
      Consequently, for the radiation of the two electrons with \(\delta=2n\pi\), a measure of imperfection of net energy conservation over time of \(\Delta\), \(I_{two}(\Delta,\alpha,\delta=2n\pi,d)\), is similarly defined as \(I(\Delta)\), which is equal to \(I(\Delta)\).

\section{Discussion of the results}According to the analysis of this paper, the time integrations of power of periodically oscillating electrons radiating into free-space over a period \(T\) are the same in CED and CWFT \cite{more33}, as shown in App. \ref{at}, regardless of the number of electrons. Therefore, for the radiation of a periodically oscillating electron or two periodically oscillating electrons, CED and CWFT dictate the same time integration of electrons' power over a period and do the same for the radiated energy flux, while net energy conservation over a period is preserved in CWFT \cite{more44}. Hence, CED and CWFT infer the same net energy conservation over a period, which corrects an analysis contradicting this \cite{madey0} and demonstrates another similarity between CED and CWFT. This further demonstrates that CWFT is not offensive to CED and thus shows another suitability of the use of CWFT as a reliable alternative theory of CED in the analysis of this paper.

For the radiation of the single electron or the two electrons with \(\delta=2n\pi\), according to \(I(\Delta)\) and \(I_{two}(\Delta,\alpha,\delta=2n\pi,d)\), it is found that although net energy conservation over time of \(\Delta\) is approximately well preserved for \(\Delta\) much greater than a period, for \(\Delta\) less than a period it is not guaranteed to be preserved. And the phasor of field with which electrons interact via the electrons' power to exchange energy is in quadrature with the resultant radiation field (the phase difference between the field on an electron and the electron's velocity is an integer multiple of \(\pi\)), which can be seen from the following relation:
\begin{widetext}
\begin{equation}\begin{split}\label{energyc0}
&\lim\limits_{R\rightarrow \infty}[W(t)-P(t+\frac{R}{c}\pm \frac{\pi}{2\omega},R)]=0,
\\&\lim\limits_{R\rightarrow \infty}[W_{two}(t,\alpha,\delta=2n\pi,d)- P_{two}(t+\frac{R}{c}\pm \frac{\pi}{2\omega},R,\alpha,\delta=2n\pi,d)]=0.\end{split}\end{equation}
\end{widetext}
This can be related to an argument of the FEL electrons' velocity-dependent amplification of light in quadrature with the acceleration-dependent radiation emitted by the electrons \cite{madey1}; the radiated energy flux exiting the infinite sized sphere is proportional to \(\lim\limits_{R\rightarrow \infty}\cos^2[\omega (t-\frac{R}{c})]\) while the electrons' retarded velocity is in phase with \(\lim\limits_{R\rightarrow \infty}[-\sin \{\omega (t-\frac{R}{c})\}]\).

For the FEL simulated, although the number of electrons can be much larger than two, the electrons' oscillations may not be perfectly coherent, the electrons' velocities are not absolutely zero, and there can be more complicated electrons' motion and thus the electrons' motion may not be pure periodic transverse oscillations, \(I(\Delta)\) and \(I_{two}(\Delta,\alpha,\delta=2n\pi,d)\) can be reasonable approximated indicator of imperfection of net energy conservation over time of \(\Delta\). The reasons for this include that the imperfection does not diminish as the number of periodically oscillating electrons increases [\(I(\Delta)=I_{two}(\Delta,\alpha,\delta=2n\pi,d)\)] and that the electrons can be approximated to be non-relativistic in the electrons' comoving frame.

Wheeler-Feynman time-symmetric theory of complete absorbing system dictates that if the causality is preserved between the electrons' power and radiated energy flux, net energy conservation over the timestep that should be much less than a period is not guaranteed to be preserved, in the simulations; if the timestep is \(0.3T\), which may not be smaller than an actual timestep, the imperfection of net energy conservation can be approximately \(50\%\). This conflicts with available simulations' strategy. Therefore, the time evolution of electrons and radiation field which is predicted by the simulations should contradict the true physics, regardless of whether in the free-electron laser the stimulated emission dominates over the spontaneous emission or not.




\begin{acknowledgements}I wish to acknowledge Professor John Madey for irreplaceable inspirations.\end{acknowledgements}

\begin{widetext}
\begin{appendix}
\section{Retarded and advanced radiation field}\label{a1}
The retarded Li{\'e}nard-Wiechert potential corresponding to one electron of position of \(\mathbf{r}_{s}(t)=z_0 e^{-i\omega t}\hat{z}\) is the following:

\begin{equation}\label{1}
    \Phi (\mathbf{r},t)_{ret}=\frac{1}{4\pi \epsilon_0} \Big\{ \frac{q}{(1-\hat{n}\cdot \beta_s)|\mathbf{r}-\mathbf{r}_s|}\Big\}_{t_{ret}},
\end{equation}

\begin{equation}\label{2}
    \mathbf{A}(\mathbf{r},t)_{ret}=\frac{\mu_0 c}{4\pi}\Big\{ \frac{q\beta_s}{(1-\hat{n}\cdot \beta_s)|\mathbf{r}-\mathbf{r}_s|} \Big\}_{t_{ret}},
\end{equation}
where \({\hat{n}}=\frac{\mathbf{r}-\mathbf{r}_s}{|\mathbf{r}-\mathbf{r}_s|}\), and \(\Phi(\mathbf{r},t)_{ret}\) becomes the following in the non-relativistic limit:

\begin{equation}
\label{3}\begin{split}
   \Phi(\mathbf{r},t)_{ret} \simeq  \frac{q}{4\pi \epsilon_0 r} \bigg\{(1+\beta_s \cos \theta)\Big(1-\frac{\delta r}{r} \Big) \bigg\}_{t_{ret}},
     \end{split}
\end{equation}
where \(\beta_s=-\frac{i\omega z_0}{c}e^{-i\omega t}\), and \(\delta r=-z_0 \cos (\omega t)\cos \theta\); \(\theta\) is the polar angle of \(\mathbf{r}\). Then, using \(t_{ret}=t-\frac{r}{c}-\frac{\delta r_{t_{ret}}}{c}\), \(\delta r_{t_{ret}}\) becomes the following:

\begin{equation}\label{4}
    \delta r_{t_{ret}}\simeq -z_0 \cos \theta \Big[ \cos \big\{\omega (t-\frac{r}{c})\big\}+\sin \big\{\omega (t-\frac{r}{c})\big\}\frac{\omega \delta r_{t_{ret}}}{c} \Big],
\end{equation}
and \(\delta r_{t_{ret}}\) becomes the following:

\begin{equation}\label{5}
    \delta r_{t_{ret}}\simeq -z_0 \cos \theta \cos \big\{\omega (t-\frac{r}{c}) \big\}.
\end{equation}
Therefore, \(\Phi(\mathbf{r},t)_{ret}\) becomes the following:

\begin{equation}\label{6}
    \Phi(\mathbf{r},t)_{ret}\simeq \frac{q}{4\pi \epsilon_0 r} \Big\{1+\frac{iU_0}{c\omega}\big(1+\frac{i}{u}\big)\cos \theta e^{-i\omega (t-\frac{r}{c})} \Big\},
\end{equation}
where \(u\equiv kr\), and \(U_0\equiv -\omega^2 z_0\), and \(\mathbf{A}(\mathbf{r},t)_{ret}\) becomes the following:

\begin{equation}\label{7}
    \mathbf{A}(\mathbf{r},t)_{ret}\simeq\frac{q\mu_0 U_0 i}{4\pi r \omega}e^{-i\omega(t-\frac{r}{c})}\hat{z}.
\end{equation}
The corresponding electromagnetic field is the following, in the spherical coordinate:
\begin{equation}\begin{split}\label{8}
   & \mathbf{E}(\mathbf{r},t)_{ret}=\frac{q}{4\pi \epsilon_0 r^2}\hat{r}+\frac{q \omega U_0}{4\pi \epsilon_0 c^3}e^{-i(\omega t-u)}\frac{1}{u}\Big\{2 \big(\frac{i}{u}-\frac{1}{u^2}\big)\cos \theta \hat{r} +\big(1+\frac{i}{u}-\frac{1}{u^2}\big)\sin \theta \hat{\theta}\Big\},
   \\&  \mathbf{B}(\mathbf{r},t)_{ret}=\frac{q\omega U_0}{4\pi \epsilon_0 c^4} e^{-i(\omega t-u)}\frac{1}{u}(1+\frac{i}{u})\sin \theta \hat{\varphi}.
\end{split}
\end{equation}
Therefore, the radiated energy flux over the boundary of a sphere of radius \(R\rightarrow \infty\) centered at the origin becomes the following:

\begin{equation}\label{ef0}\begin{split}
    \lim\limits_{R\rightarrow \infty}P(t,R)&\equiv\lim\limits_{R\rightarrow \infty}\frac{1}{\mu_0}\int (Re[\mathbf{E}_{ret}]\times Re[\mathbf{B}_{ret}])\cdot \hat{r}ds\\&=\lim\limits_{R\rightarrow \infty}\frac{q^2U_0^2}{6\pi \epsilon_0 c^3}\Big[\cos^2\{\omega(t-\frac{R}{c})\}+\frac{\sin \{2\omega(t-\frac{R}{c})\}}{kR}-\frac{\cos \{2\omega(t-\frac{R}{c})\}}{(kR)^2}-\frac{\sin \{2\omega(t-\frac{R}{c})\}}{2(kR)^3}\Big]
    \\&=\lim\limits_{R\rightarrow \infty}\frac{q^2U_0^2}{6\pi \epsilon_0 c^3}\cos^2\{\omega(t-\frac{R}{c})\}.
\end{split}\end{equation}

The advanced Li{\'e}nard-Wiechert potential corresponding to the periodically oscillating electron is the following:
\begin{equation}\label{18}
     \Phi (\mathbf{r},t)_{adv}=\frac{1}{4\pi \epsilon_0} \Big\{ \frac{q}{(1+{\hat{n}}\cdot {\beta}_s)|\mathbf{r}-\mathbf{r}_s|}\Big\}_{t_{adv}},
\end{equation}

\begin{equation}\label{19}
      \mathbf{A}(\mathbf{r},t)_{adv}=\frac{\mu_0 c}{4\pi}\Big\{ \frac{q{\beta}_s}{(1+{\hat{n}}\cdot {\beta}_s)|\mathbf{r}-\mathbf{r}_s|} \Big\}_{t_{adv}}.
\end{equation}
Then, using

\begin{equation}\label{20}
\delta r_{t_{adv}}\simeq -z_0 \cos \theta \cos \{\omega (t+\frac{r}{c})\},
\end{equation}
\(\Phi (\mathbf{r},t)_{adv}\) becomes the following:

\begin{equation}\label{21}
    \Phi (\mathbf{r},t)_{adv}\simeq \frac{q}{4\pi \epsilon_0 r} \Big\{1+\frac{iU_0}{c\omega}\big(-1+\frac{i}{u}\big)\cos \theta e^{-i\omega (t+\frac{r}{c})} \Big\},
\end{equation}
and \( \mathbf{A}(\mathbf{r},t)_{adv}\) becomes the following:

\begin{equation}\label{22}
     \mathbf{A}(\mathbf{r},t)_{adv}\simeq\frac{q\mu_0 U_0 i}{4\pi r \omega}e^{-i\omega(t+\frac{r}{c})}\hat{z}.
\end{equation}
Then, the advanced electromagnetic field is the following:

\begin{equation}\begin{split}\label{23}
    &\mathbf{E}(\mathbf{r},t)_{adv}=\frac{q}{4\pi \epsilon_0 r^2}\hat{r}+\frac{q \omega U_0}{4\pi \epsilon_0 c^3}e^{-i(\omega t+u)}\frac{1}{u}\Big\{2 \big(-\frac{i}{u}-\frac{1}{u^2}\big)\cos \theta \hat{r} +\big(1-\frac{i}{u}-\frac{1}{u^2}\big)\sin \theta \hat{\theta}\Big\},
    \\& \mathbf{B}(\mathbf{r},t)_{adv}=\frac{q\omega U_0}{4\pi \epsilon_0 c^4} e^{-i(\omega t+u)}\frac{1}{u}(-1+\frac{i}{u})\sin \theta \hat{\varphi}.
\end{split}
\end{equation}

\section{\(\mathbf{E}_{abs}\) of Wheeler-Feynman time-symmetric theory of complete absorbing system}\label{bb}

\(\mathbf{E}_{abs,\perp}(\mathbf{r},t)\) [portion of \(\mathbf{E}_{abs}(\mathbf{r},t)\) perpendicular to \(\mathbf{U}(t)\)] is the following in the coordinate of which \(\hat{z}\) that is along \(\mathbf{r}\), according to CWFT:

 \begin{equation}\begin{split}\label{14}
    &\mathbf{E}_{abs,\perp}(\mathbf{r},t)=-\frac{i\omega q}{4\pi \epsilon_0 c^3}U_0 e^{-i \omega t}\int_{-1}^{1} \frac{1}{2}d(\cos(\mathbf{r},\mathbf{r}_k))\int_{0}^{2\pi} \frac{d\phi_d}{2\pi}\cos(\mathbf{r}_k,\mathbf{U}(t))e^{iu\cos(\mathbf{r},\mathbf{r}_k)}\times \\&\Big[\{-\sin(\mathbf{r},\mathbf{r}_k)\cos\phi'+\cos(\mathbf{r}_k,\mathbf{U}(t))\sin(\mathbf{r},\mathbf{U}(t))\cos \phi\}\hat{x}+\{-\sin(\mathbf{r},\mathbf{r}_k)\sin \phi' +\cos(\mathbf{r}_k,\mathbf{U}(t))\sin(\mathbf{r},\mathbf{U}(t))\sin \phi\}\hat{y}\\&+\{-\cos(\mathbf{r},\mathbf{r}_k)+\cos(\mathbf{r}_k,\mathbf{U}(t))\cos(\mathbf{r},\mathbf{U}(t))\}\hat{z}\Big];
 \end{split}\end{equation}
\(\mathbf{r}_k\) is location of the particle in the absorber, \(\phi\) and \(\phi'\) are the azimuthal angle of \(\mathbf{U}(t)\) and \(\mathbf{r}_k\), respectively, and \(\phi_d\) is dihedral angle between the \((\mathbf{r},\mathbf{U})\) and \((\mathbf{r},\mathbf{r}_k)\) planes. The \(x\)-component of \(\mathbf{E}_{abs,\perp}(\mathbf{r},t)\) of Eqn. (\ref{14}) becomes the following [\(\theta\) and \(\theta'\) are polar coordinates of \(\mathbf{U}(t)\) and \(\mathbf{r}_k\), respectively]:

\begin{equation}\label{x}\begin{split}
   E_{{abs,\perp}_x}(\mathbf{r},t)&=\frac{-i\omega q}{4\pi \epsilon_0 c^3}U_0 e^{-i \omega t}\int_{-1}^{1} \frac{1}{2}d(\cos \theta')\int_{0}^{2\pi} \frac{d\phi'}{2\pi}\cos(\mathbf{r}_k,\mathbf{U}(t))e^{iu\cos\theta'}\{-\sin \theta' \cos\phi'+\cos(\mathbf{r}_k,\mathbf{U}(t))\sin\theta \cos \phi\}.
    \end{split}
\end{equation}
The first term of \(E_{{abs,\perp}_x}(\mathbf{r},t)\) becomes the following:

\begin{equation}\label{x1}\begin{split}
&\frac{i\omega q}{4\pi \epsilon_0 c^3}U_0 e^{-i \omega t}\int_{-1}^{1} \frac{1}{2}d(\cos \theta')\int_{0}^{2\pi} \frac{d\phi'}{2\pi}\{\sin \theta \sin \theta' \cos(\phi-\phi')+\cos \theta \cos \theta'\}e^{iu\cos\theta'}\sin \theta' \cos\phi'
\\&=\frac{i\omega q}{16\pi \epsilon_0 c^3}U_0 e^{-i \omega t}C_1\sin \theta \cos \phi,
\end{split}\end{equation}
where \(C_1\equiv \int_{-1}^{1}(1-x^2)e^{iux}dx=4\frac{\sin u-u\cos u}{u^3}\), and the second term of \(E_{{abs,\perp}_x}\) becomes the following:

\begin{equation}\label{x2}
-\frac{i\omega q}{16\pi \epsilon_0 c^3}U_0 e^{-i \omega t}C_2 \sin \theta \cos \phi,    \end{equation}
where \(C_2\equiv \frac{1}{\pi}\int_{-1}^{1}d(\cos \theta')\int_{0}^{2\pi}d\phi' \cos^2 (\mathbf{r_k},\mathbf{U}(t))e^{iu\cos \theta'}\).
Hence, \(E_{{abs,\perp}_x}(\mathbf{r},t)\) becomes the following:

\begin{equation}\label{xf}
    E_{{abs,\perp}_x}(\mathbf{r},t)=\frac{i\omega q}{16\pi \epsilon_0 c^3}U_0 e^{-i \omega t}(C_1-C_2)\sin \theta \cos \phi.
\end{equation}
The \(y\)-component of \(\mathbf{E}_{abs,\perp}(\mathbf{r},t)\) becomes the following:

\begin{equation}\begin{split}\label{y}
 E_{{abs,\perp}_y}(\mathbf{r},t)&=-\frac{i\omega q}{4\pi \epsilon_0 c^3}U_0 e^{-i \omega t}
 \int_{-1}^{1} \frac{1}{2}d(\cos \theta')\int_{0}^{2\pi} \frac{d\phi'}{2\pi}\cos(\mathbf{r}_k,\mathbf{U}(t))e^{iu\cos\theta'}\{-\sin \theta' \sin\phi'+\cos(\mathbf{r}_k,\mathbf{U}(t))\sin\theta \sin \phi\}.    
\end{split}\end{equation}
The first term of \(E_{{abs,\perp}_y}(\mathbf{r},t)\) becomes the following:

\begin{equation}\label{y1}\begin{split}
&\frac{i\omega q}{4\pi \epsilon_0 c^3}U_0 e^{-i \omega t}\int_{-1}^{1} \frac{1}{2}d(\cos \theta')\int_{0}^{2\pi} \frac{d\phi'}{2\pi}\{\sin \theta \sin \theta' \cos(\phi-\phi')+\cos \theta \cos \theta'\}e^{iu\cos\theta'}\sin \theta' \sin\phi'
\\&=\frac{i\omega q}{16\pi \epsilon_0 c^3}U_0 e^{-i \omega t}C_1\sin \theta \sin \phi,
\end{split}\end{equation}
and the second term of \(E_{{abs,\perp}_y}(\mathbf{r},t)\) is the following:

\begin{equation}\label{y2}
    -\frac{i\omega q}{16\pi \epsilon_0 c^3}U_0 e^{-i \omega t}C_2 \sin \theta \sin \phi.    
\end{equation}
Therefore, \(E_{{abs,\perp}_y}(\mathbf{r},t)\) becomes the following:

\begin{equation}\label{yf}
      E_{{abs,\perp}_y}=\frac{i\omega q}{16\pi \epsilon_0 c^3}U_0 e^{-i \omega t}(C_1-C_2)\sin \theta \sin \phi.
\end{equation}
The \(z\)-component of \(\mathbf{E}_{abs,\perp}(\mathbf{r},t)\) is the following:

\begin{equation}\begin{split}\label{z}
   E_{{abs,\perp}_z}(\mathbf{r},t)=-\frac{i\omega q}{4\pi \epsilon_0 c^3}U_0 e^{-i \omega t}\int_{-1}^{1} \frac{1}{2}d(\cos \theta')\int_{0}^{2\pi} \frac{d\phi'}{2\pi}\cos(\mathbf{r}_k,\mathbf{U}(t))e^{iu\cos\theta'}\{-\cos \theta' +\cos(\mathbf{r}_k,\mathbf{U}(t))\cos\theta\}.
\end{split}
\end{equation}
The first term of \(E_{{abs,\perp}_z}(\mathbf{r},t)\) is the following:

\begin{equation}\begin{split}\label{z1}
   & \frac{i\omega q}{4\pi \epsilon_0 c^3}U_0 e^{-i \omega t}\int_{-1}^{1} \frac{1}{2}d(\cos \theta')\int_{0}^{2\pi} \frac{d\phi'}{2\pi}\{\sin \theta \sin \theta'\cos(\phi-\phi')+\cos \theta \cos \theta'\}e^{iu\cos\theta'}\cos \theta'
   \\&= \frac{i\omega q}{8\pi \epsilon_0 c^3}U_0 e^{-i \omega t}C_3 \cos \theta, 
\end{split}\end{equation}
where \(C_3\equiv \int_{-1}^{1} x^2 e^{iux}dx=2\frac{(u^2-2)\sin u+2u\cos u}{u^3}\). And the second term of \(E_{{abs,\perp}_z}(\mathbf{r},t)\) is the following:

\begin{equation}\label{z2}
     -\frac{i\omega q}{16\pi \epsilon_0 c^3}U_0 e^{-i \omega t}C_2 \cos \theta.
\end{equation}
Therefore, \(E_{{abs,\perp}_z}(\mathbf{r},t)\) becomes the following:

\begin{equation}\label{zf}
     \frac{i\omega q}{16\pi \epsilon_0 c^3}U_0 e^{-i \omega t}(2C_3-C_2) \cos \theta.
\end{equation}
Hence, the resultant \(\mathbf{E}_{abs,\perp}(\mathbf{r},t)\) becomes the following:

\begin{equation}\label{eperp}
    \mathbf{E}_{abs,\perp}(\mathbf{r},t)=  \frac{i\omega q}{16\pi \epsilon_0 c^3}U_0 e^{-i \omega t} \{(C_1-C_2)\sin \theta (\cos \phi \hat{x}+\sin \phi \hat{y})+(2C_3-C_2)\cos \theta \hat{z}\}.
\end{equation}
To express \(\mathbf{E}_{abs,\perp}(\mathbf{r},t)\) in a coordinate of which \(\hat{z}\) that is along \(\mathbf{r}_s(t)\) (in the remainder of this App. \ref{bb}, this coordinate system is used), Eqn. (\ref{eperp}) should be transformed by the following matrix:

\[\begin{pmatrix} 
    \cos \theta \cos \phi & \cos \theta \sin \phi &-\sin \theta \\
    -\sin \phi & \cos \phi & 0 \\
    \sin \theta \cos \phi & \sin \theta \sin \phi &\cos \theta
\end{pmatrix}=
\begin{pmatrix} 
\cos \theta & 0 &-\sin \theta \\
0 & 1 & 0\\
\sin \theta & 0 & \cos \theta
\end{pmatrix}
\begin{pmatrix}
\cos \phi & \sin \phi & 0\\
-\sin \phi & \cos \phi &0\\
0&0&1
\end{pmatrix}
\]
The \(-x\)-component of the transformed \(\mathbf{E}_{abs,\perp}(\mathbf{r},t)\) is the \(\hat{\rho}\) (axial unit vector in the cylindrical coordinate)-component in the new coordinate, while \(y\) and \(z\)-components vanish. The resultant \(\mathbf{E}_{abs,\perp}(\mathbf{r},t)\) becomes the following:

\begin{equation}\begin{split}\label{15}
    \mathbf{E}_{abs,\perp}(\mathbf{r},t)=\frac{i\omega q}{16\pi \epsilon_0 c^3}U_0 e^{-i \omega t}(2C_3-C_1)\sin \theta \cos \theta \hat{\rho}=\frac{q\omega U_0}{4\pi \epsilon_0 c^3}e^{-i\omega t}\frac{1}{u}\Big[i\sin u+i\frac{3}{u}\cos u-i\frac{3}{u^2}\sin u\Big]\sin \theta \cos \theta \hat{\rho},
\end{split}\end{equation}
and \(\mathbf{E}_{abs,\parallel}\) of CWFT is given as the following:

\begin{equation}\label{16}
    \mathbf{E}_{abs,\parallel}(\mathbf{r},t)= \frac{q\omega U_0}{4\pi \epsilon_0 c^3} e^{-i\omega t}\frac{1}{u}\Big\{-i\sin u \sin^2 \theta +i\big(\frac{\cos u}{u}-\frac{\sin u}{u^2}\big)(3\cos^2 \theta-1)\Big\}\hat{z}.
\end{equation}
Therefore, \(\mathbf{E}_{abs}(\mathbf{r},t)=\mathbf{E}_{abs,\parallel}(\mathbf{r},t)+\mathbf{E}_{abs,\perp}(\mathbf{r},t)\) becomes the following:

\begin{equation}\begin{split}\label{17}
    &\mathbf{E}_{abs}(\mathbf{r},t)\\&= \frac{q\omega U_0}{4\pi \epsilon_0 c^3}\frac{e^{-i\omega t}}{ u}\Big[\bigg\{\big(\frac{i}{u}-\frac{1}{u^2}\big)e^{iu}+\big(\frac{i}{u}+\frac{1}{u^2}\big)e^{-iu}\bigg\}\cos \theta \hat{r}
   +\frac{1}{2}\bigg\{\big(1+\frac{i}{u}-\frac{1}{u^2}\big)e^{iu}+\big(-1+\frac{i}{u}+\frac{1}{u^2}\big)e^{-iu}\bigg\}\sin \theta \hat{\theta}\Big].
\end{split}\end{equation}

\section{Two radiating electrons' power and radiated energy flux}\label{td}

\(W_{two}(t,\alpha,\delta,d)\) can be computed as the following for \(d\) of order lower than \(\mathcal{O}\Big(\frac{U_0}{c\omega}\Big)\):

\begin{equation}\begin{split}\label{ft}
     &W_{two}(t,\alpha,\delta,d)=\frac{q^2U_0^2}{6\pi \epsilon_0 c^3}\{1-\cos \delta \cos (2\omega t -\delta)\}-q\Big[Re[\mathbf{E}_{ret,2}(\mathbf{r}_1,t)]\cdot Re[\dot{\mathbf{r}}_1]+Re[\mathbf{E}_{ret,1}(\mathbf{r}_2,t)]\cdot Re[\dot{\mathbf{r}}_2]\Big]
     \\&=\frac{q^2U_0^2}{6\pi \epsilon_0 c^3}\{1-\cos \delta \cos (2\omega t -\delta)\}+\frac{q^2 U_0}{2\pi \epsilon_0 d^2 \omega}\sin (\frac{\delta}{2})\cos \alpha \cos(\omega t-\frac{\delta}{2})\\&-\frac{q^2U_0^2}{4\pi \epsilon_0 c^3}\Big[Re[A]\sin(2\omega t-\delta)-Im[A]\cos (2\omega t-\delta)+\cos \delta Im[A]\Big],
\end{split}
    \end{equation}
   where 
\begin{equation}\begin{split}\label{A}
    A\equiv \frac{e^{ikd}}{kd} \Big[2\Big\{\frac{i}{kd}-\frac{1}{(kd)^2}\Big\}\cos ^2 \alpha-\Big\{1+\frac{i}{kd}-\frac{1}{(kd)^2}\Big\}\sin^2 \alpha\Big].
\end{split}\end{equation}
Therefore, Eqn. (\ref{ft}) becomes the following:

\begin{equation}\begin{split}\label{wf}    W_{two}(t,\alpha,\delta,d)=\frac{q^2U_0^2}{6\pi \epsilon_0 c^3} \big\{1-\cos \delta \cos (2\omega t-\delta)\big\}.\end{split}    \end{equation}
       For \(d\) of order of \(\mathcal{O}\Big(\frac{U_0}{c\omega}\Big)\), \(W_{two}(t,\alpha,\delta,d)\) should be equal to Eqn. (\ref{ft}) without the term containing \(Re[A]\), which becomes the following:
 
    \begin{equation}\begin{split}\label{l5}
    &W_{two}(t,\alpha,\delta,d)=\frac{q^2U_0^2}{6\pi \epsilon_0 c^3} \Big[\big\{1-\cos \delta \cos (2\omega t-\delta)\big\}    +3\big\{\cos \delta -\cos (2\omega t-\delta)\}X(d,\alpha)         \Big],
    \end{split}\end{equation}
    where \(X(d,\alpha)\) is defined below:
    \begin{equation}\label{aas}
X(d,\alpha)\equiv\Big[\cos^2 \alpha \big\{-\frac{\cos (kd)}{(kd)^2}+\frac{\sin (kd)}{(kd)^3}\big\}+\frac{\sin^2 \alpha}{2}\Big\{\frac{\sin (kd)}{kd}+\frac{\cos (kd)}{(kd)^2}-\frac{\sin (kd)}{(kd)^3}\Big\}\Big].
\end{equation} The radiated energy flux over the boundary of sphere of radius \(R\) centered at the origin is the following:

\begin{equation}\label{n00}
 P_{two}(t,R,\alpha,\delta,d)=\frac{1}{\mu_0}\int \Big[Re[\mathbf{E}_{ret,1}+\mathbf{E}_{ret,2}]\times Re[ \mathbf{B}_{ret,1}+\mathbf{B}_{ret,2}] \Big]\cdot \hat{r} da.
 \end{equation}
Then, \(\lim\limits_{R\rightarrow \infty}P_{two}(t,R,\alpha,\delta,d)\) is the following [\(\Delta u\equiv \frac{kd}{2}(\sin \theta \cos \phi \sin \alpha+\cos \theta \cos \alpha)\), and \(\theta\) and \(\phi\) are the polar and azimuthal angles of the spherical coordinate, respectively]:

\begin{equation}\begin{split}\label{n1}
   &\lim\limits_{R\rightarrow \infty}P_{two}(t,R,\alpha,\delta,d)
    \\&=\lim\limits_{R\rightarrow \infty}\frac{q^2 U_0^2}{32\pi^2 \epsilon_0 c^3} \int_{\phi=0}^{\phi=2\pi} \int_{\theta=0}^{\theta=\pi}\sin^3 \theta\Big\{Re[e^{-2i\omega(t-\frac{R}{c})}\{e^{i\Delta u}+e^{i(-\Delta u +\delta)}\}^2]+4\cos^2 \big(\Delta u-\frac{\delta}{2}\big)\Big\}d\phi d\theta
    \\&=\lim\limits_{R\rightarrow \infty}\frac{q^2 U_0^2}{4\pi^2 \epsilon_0 c^3} \cos^2 \{\omega(t-\frac{R}{c})-\frac{\delta}{2}\} \int_{\phi=0}^{\phi=2\pi} \int_{\theta=0}^{\theta=\pi}\sin^3 \theta\cos^2(\Delta u-\frac{\delta}{2})d\phi d\theta.
   \end{split}
\end{equation}
The above integral expression can be calculated as the following [\(\text{J}_n(x)\) is the Bessel functions of the first kind, and \(C_n^{\nu}(x)\) is the Gegenbauer polynomials]:

\begin{equation}\begin{split}\label{int}
    &\int_{\phi=0}^{\phi=2\pi} \int_{\theta=0}^{\theta=\pi}\sin^3 \theta\cos^2(\Delta u-\frac{\delta}{2})d\phi d\theta
   =\pi \int_{0}^{\pi}\sin^3 \theta \{1+  \text{J}_0 (kd\sin \alpha \sin \theta)\cos (kd \cos \alpha \cos \theta -\delta)\}d\theta
    \\&=\frac{4}{3}\pi \Big[1+\cos \delta \int_{0}^{\frac{\pi}{2}}\sin \theta \{C_0^\frac{1}{2}(\cos \theta)-C_2^\frac{1}{2}(\cos \theta)\} \cos (kd \cos \alpha \cos \theta)\text{J}_0 (kd\sin \alpha \sin \theta)d\theta \Big]
    \\&=\frac{4}{3}\pi \Big[1+\cos \delta \sqrt{\frac{\pi}{2kd}}\Big\{\frac{1}{2}(3\cos^2 \alpha -1)\text{J}_{\frac{5}{2}}(kd)+\text{J}_{\frac{1}{2}}(kd)\Big\} \Big]
    =\frac{4}{3}\pi \big[1+3X(d,\alpha)\cos \delta   \big],
\end{split}
\end{equation}
where the \(\theta\)-integration is given in Ref. \cite{int}.
Therefore, \(\lim\limits_{R\rightarrow \infty}P_{two}(t,R,\alpha,\delta,d)\) becomes the following:

\begin{equation}\begin{split}\label{ef}
    \lim\limits_{R\rightarrow \infty}P_{two}(t,R,\alpha,\delta,d)=&\lim\limits_{R\rightarrow \infty}\frac{q^2 U_0^2}{3\pi \epsilon_0 c^3} \cos^2 \{\omega(t-\frac{R}{c})-\frac{\delta}{2}\}
     \big[1+3X(d,\alpha)\cos \delta   \big].
\end{split}
    \end{equation}

\section{The time integration of the electrons' power over a period}\label{at}

In CED, for radiation of the single electron, the field at the electron can be decomposed as the following:

\begin{equation}\begin{split}\label{dv}
    \lim\limits_{\mathbf{r}\rightarrow \mathbf{r}_s} \mathbf{E}_{ret}(\mathbf{r})=\lim\limits_{\mathbf{r}\rightarrow \mathbf{r}_s}[ \mathbf{E}_{abs}(\mathbf{r})+\frac{1}{2}\{\mathbf{E}_{ret}(\mathbf{r})+\mathbf{E}_{adv}(\mathbf{r})\}].
\end{split}\end{equation}
As \(\lim\limits_{\mathbf{r}\rightarrow \mathbf{r}_s} \frac{1}{2}\{\mathbf{E}_{ret}(\mathbf{r})+\mathbf{E}_{adv}(\mathbf{r})\}\) is divergent, and its direction is indeterminate, the electron's power is indeterminate in CED. However, as

\begin{equation}\label{eq}
-\lim\limits_{\mathbf{r}\rightarrow \mathbf{r}_s}\int_{t_0}^{t_0+T}q\frac{1}{2}Re[\mathbf{E}_{ret}(\mathbf{r})+\mathbf{E}_{adv}(\mathbf{r})] \cdot Re[\dot{\mathbf{r}}_s(t)] dt=0,
\end{equation}
the following relation can be obtained according to \(W(t)\) of Eqn. (\ref{ew}) and Eqn. (\ref{dv}):

\begin{equation}\label{equal}
    \int_{t_0}^{t_0+T} W(t)dt=-\lim\limits_{\mathbf{r}\rightarrow \mathbf{r}_s}\int_{t_0}^{t_0+T}qRe[\mathbf{E}_{ret}(\mathbf{r})] \cdot Re[\dot{\mathbf{r}}_s(t)] dt,
\end{equation}
which means that the time integrations of the electron's power over a period is the same in CED and CWFT
. Then, according to \(\mathbf{E}^{(i)}(t)\) of Eqs. (\ref{e1s2})-(\ref{p1}) and the following relation,

\begin{equation}\label{p2}
\int_{t_0}^{t_0+T}\frac{1}{2}\{Re[\mathbf{E}_{ret,2}(\mathbf{r}_1,t)+\mathbf{E}_{adv,2}(\mathbf{r}_1,t)]\cdot Re[\dot{\mathbf{r}}_1(t)]+Re[\mathbf{E}_{ret,1}(\mathbf{r}_2,t)+\mathbf{E}_{adv,1}(\mathbf{r}_2,t)]\cdot Re[\dot{\mathbf{r}}_2(t)]\}dt=0,
\end{equation}
the time integrations of the power of periodically oscillating electrons radiating into free-space over a period is shown to be the same in CED and CWFT, regardless of the number of electrons.

\end{appendix}
\end{widetext}


\begin{thebibliography}{99}
\bibitem{more33}Although the electrons' power is indeterminate in classical electrodynamics, by doing the time integration prior to the limiting process of the field approaching to the electrons, the integration can be computed.
\bibitem{more44}For the two electrons' radiation, \(\int_{t'_0}^{t'_0+T}W_{two}(t,\alpha,\delta,d)dt\) is equal to \(\int_{t_0}^{t_0+T}\lim\limits_{R\rightarrow \infty}P_{two}(t,R,\alpha,\delta,d)dt\).
\end{thebibliography}
\end{document}